\documentclass[aps,twocolumn,groupedaddress]{revtex4-1}
\usepackage[dvips]{graphicx}

\begin{document}
\title{Field Induced Quantum Phase Transitions in $S=1/2$ $J_1$-$J_2$  Heisenberg Model on the Square Lattice}
\author{Katsuhiro Morita}
\email[e-mail:]{morita@cmpt.phys.tohoku.ac.jp}
\author{Naokazu Shibata}
\affiliation{Department of Physics, Tohoku University, Aoba-ku, Sendai 980-8578 Japan}

\date{\today}
\begin{abstract}
We study the magnetic field dependence of the ground state of the $S=1/2$ $J_1$-$J_2$ Heisenberg model on the square lattice by the density matrix renormalization group (DMRG) method. With the use of the sine-square deformation, we obtain eight different ground states including plaquette valence-bond crystal with a finite spin gap, transverse N$\acute{\rm e}$el, transverse stripe, 1/2 magnetization plateau with up-up-up-down (uuud), and three new states we named the Y-like, V-like, and $\Psi$ states around $J_2/J_1$ =0.55--0.6. The phase transitions from the transverse N$\acute{\rm e}$el (at $J_2/J_1$ = 0.55) and stripe (at $J_2/J_1$ = 0.6) states to the uuud and Y-like states, respectively, are discontinuous, as in the case of a spin flop.
\end{abstract}
\maketitle
\section{Introduction}
Frustrated quantum spin systems exhibit exotic ground states such as valence bond solid (VBS), spin-nematic, and quantum spin liquid (QSL) states with the effects of quantum fluctuations and frustration \cite{Mila,Balents}.
Even in classical systems, frustration generally leads to a large number of degenerate low-energy states, and thermal fluctuation sometimes brings about qualitatively new states \cite{classical1,classical2,classical3}. 
The $J_1$-$J_2$ square lattice is such a two-dimensional frustrated system \cite{exp1,exp2,exp3}. 
Its classical ground state for $J_2/J_1$ $<$ 0.5 is a simple N$\acute{\rm e}$el state, whereas it is reconstructed to form two $\sqrt{2}$ $\times\!\sqrt{2}$ sublattice N$\acute{\rm e}$el states for $J_2/J_1$ $>$ 0.5.
The degeneracy originating from the relative direction of the two sublattice N$\acute{\rm e}$el states is solved at a finite temperature by thermal fluctuations with a collinear stripe order \cite{sqclass1,sqclass2}.
At $J_2/J_1$ $=$ 0.5, the classical ground state has macroscopic degeneracy and the thermodynamic properties are highly non-trivial.
In the $S$ = 1/2 quantum spin system, several novel ground states such as plaquette valence-bond crystal (PVBC), columnar valence-bond crystal (VBC), and spin liquids with or without a spin gap have been predicted by numerical studies such as density matrix renormalization group (DMRG) calculations \cite{DMRG1,DMRG2}, exact diagonalizations \cite{ED1,ED2}, and other numerical simulations \cite{FM,other1,other2,other3,other4,other5,other6}, but the true nature of the ground state has not been determined, and no consensus exists yet.

In a magnetic field, the presence of a 1/2 magnetization plateau of full magnetization with a up-up-up-down (uuud) structure has been predicted in addition to the transverse N$\acute{\rm e}$el and stripe phases by analysis involving exact diagonalizations \cite{uuud1,uuud2} and linear spin wave theory \cite{uuud3}.
However, the exact diagonalization studies were limited to small systems of up to about 36 spins, and the spin wave analysis assumed only three magnetic structures.
Since the correct low-energy quantum fluctuations are reproduced only in sufficiently large systems and the possible magnetic structures are not limited to the three states, the possibility of finding new quantum states in magnetic fields still exists.

\section{Model and Method}
In this paper, we examine the $S$=1/2 $J_1$-$J_2$ Heisenberg model on the square lattice in a magnetic field by performing large-scale DMRG calculations.
The Hamiltonian of this model is defined as

\begin{figure}
\begin{center}
\includegraphics[width=30mm]{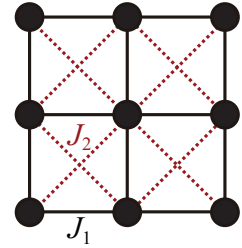}
\caption{(Color online) $J_1$-$J_2$ square lattice. $J_1$ and $J_2$ represent nearest- and next-nearest-neighbor exchange interactions, respectively.
\label{J1-J2-lattice}}
\end{center}
\end{figure}
\begin{figure}
\begin{center}
\includegraphics[width=64mm]{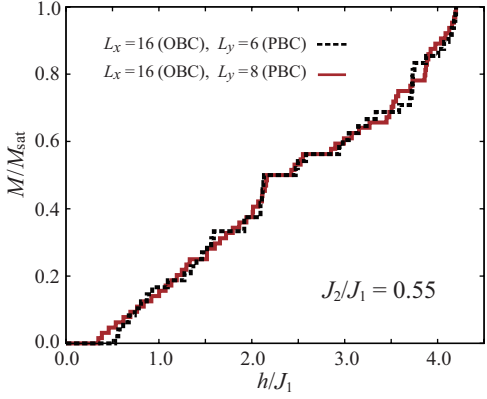}
\caption{(Color online) $M/M_{\rm sat}$ vs $h/J_1$ in cylindrical systems of $L_x$ = 16 at $J_2/J_1$ = 0.55.
The red solid and black dashed lines correspond to $L_y$ = 8 and 6, respectively.
\label{simpleM-H}}
\end{center}
\end{figure}

 

\begin{eqnarray} 
H &=& J_1\sum_{nn} \mathbf{S}_i \cdot \mathbf{S}_j
+ J_2\sum_{nnn} \mathbf{S}_i\cdot \mathbf{S}_j - h\sum_i S^{z}_i,
\end{eqnarray}
where $nn$ and $nnn$ represent nearest- and next-nearest-neighbor pairs, respectively (see Fig.~\ref{J1-J2-lattice}). 
In the present study, we focus on the strongly frustrated states of the model and calculate the ground state at $J_2/J_1$ = 0.45, 0.55, 0.6, and 0.675.
Before we analyze the ground state of this model, we first show the magnetization processes of finite cylinders of length $L_x = 16$ with open-boundary conditions (OBC) and $L_y = 6$ and 8 with periodic-boundary conditions (PBC).
The results obtained by the DMRG method are shown in Fig.~\ref{simpleM-H}, where the position of the magnetization plateaus strongly depends on the circumference of the cylinder, $L_y$.
\begin{figure}
\begin{center}
\includegraphics[width=60mm]{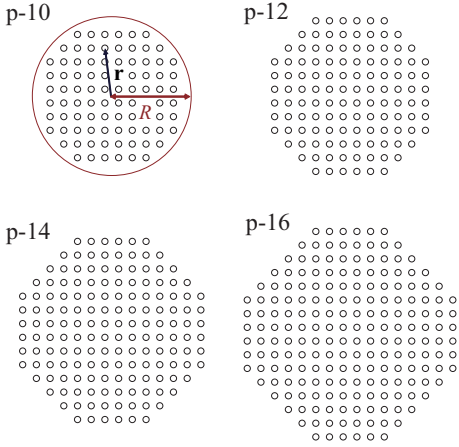}
\caption{(Color online) Four octagonal clusters used in the present calculation by the grand canonical SSD method.
The position vector from the center and the radius of the system boundary are represented by $\mathbf{r}$ and $R$, respectively. 
\label{ssdcp}}
\end{center}
\end{figure} 
This size dependence is mainly caused by the periodic boundary conditions on $L_y$, and we need to perform systematic calculations for large $L_y$ beyond the correlation length of the ground state to confirm the presence of the plateaus.
We also need to remove the effect of open-boundary conditions on $L_x$, where edge spins contribute to an artifactual shift of the plateaus.
For these reasons, it is not easy to obtain true bulk properties from finite systems of available sizes under usual boundary conditions.
To overcome this difficulty, we carry out ground canonical analysis with the recently developed sine-square deformation (SSD) method \cite{ssd1,ssd2,ssd3,ssd4}.
The SSD deforms the original Hamiltonian of Eq. (1) to that locally rescaled by the function $f(\mathbf{r})$ as
\begin{eqnarray} 
H &=& J_1\sum_{nn}f\left(\frac{\mathbf{r}_i+\mathbf{r}_j}{2}\right) \mathbf{S}_i \cdot \mathbf{S}_j \nonumber \\
   &+& J_2\sum_{nnn}f\left(\frac{\mathbf{r}_i+\mathbf{r}_j}{2}\right) \mathbf{S}_i\cdot \mathbf{S}_j 
   - h\sum_i f(\mathbf{r}_i)S^{z}_i,
\end{eqnarray}
where $f(\mathbf{r})$ is a decreasing function of $|\mathbf{r}|$ from the center of the system defined by $f\left(\mathbf{r}\right) = \frac{1}{2}\left[1+\rm{cos}\left(\frac{\pi |\mathbf{r}|}{\it{R}}\right)\right]$, which vanishes at the boundary of the system at radius $R$ as shown in Fig.~\ref{ssdcp}.
\begin{figure}
\begin{center}
\includegraphics[width=64mm]{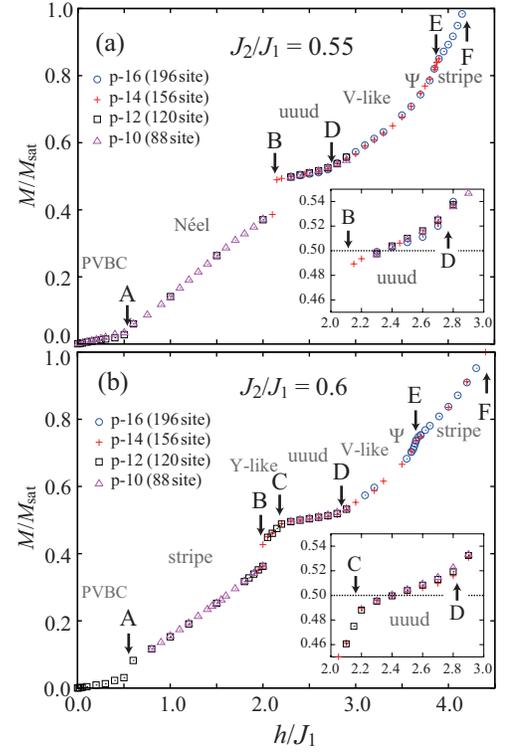}
\caption{(Color online) $M/M_{\rm sat}$ vs $h/J_1$ at $J_2/J_1$ = 0.55 (a) and 0.6 (b) obtained by the grand canonical SSD method. Arrows show the positions of anomalies in the magnetization process, indicating the presence of field-induced phase transitions.  The inset shows the size dependence of the magnetization $M/M_{\rm sat}$ around the 1/2 plateau. Notice the change in the size dependence and the slope at the transitions.
\label{ssdM-H}}
\end{center}
\end{figure}
The interactions of the spins near the boundaries with the main part of the system with a small energy scale efficiently suppress the artifactual oscillations of physical quantities induced at the edge of the system and act as a buffer of magnetic moments to maintain the optimal magnetization of the main system. 
This technique is known to reduce finite-size effects and reasonably reproduce correct bulk properties \cite{ssd4}.
In the present study, we use the four octagon clusters p-$n$ shown in Fig.~\ref{ssdcp}, where $n$ is the largest number of spins aligned in one direction. 
The small clusters (p-10 and p-12) are used in low magnetic fields, where large quantum entanglement entropy appears in the ground state.
Since the accuracy of the DMRG calculations is improved in a higher magnetic field with weak entanglement entropy, the large clusters (p-14 and p-16) are used in this case. In the present DMRG calculations, the number of states, $m$, maintained in each block is 500 -- 10000.
The truncation error is around $2.0\times 10^{-5}$ at low fields and less than $5.0\times 10^{-6}$ in other cases.
The magnetizations and energies of the systems are evaluated using the central 6 $\times$ 6 sites in the original octagonal clusters.

\section{Results}
\subsection{Magnetization process}
Figure~\ref{ssdM-H} presents the magnetization $M/M_{\rm sat}$ ($M_{\rm sat}$ is the saturated magnetization) at $J_2/J_1$ = 0.55 and 0.6.
We find several anomalies in the magnetization curves that indicate field-induced magnetic phase transitions.
The magnetization plateau at $M/M_{\rm sat}$ = 1/2 shown in Fig.~\ref{ssdM-H} has a shallow slope that is characteristic of finite-size systems in the present calculation \cite{ssd4}. With increasing system size, the slope in the plateau decreases, as shown in the insets of Fig.~\ref{ssdM-H}, and the slope seems to vanish in large systems.
As shown in Fig.~\ref{ssdM-H} by arrows, clear anomalies, such as a sudden change in the slope and a jump in the magnetization curve, are found. 
The detailed changes in the magnetic structure of the phases across these anomalies are analyzed from the real-space structures of the local magnetization $\langle S^z_i\rangle$ and the nearest- and next-nearest-neighbor correlation functions 
$\langle\mathbf{S}_i\cdot\mathbf{S}_j\rangle$ or $\langle S^+_iS^-_j\rangle$. 

\begin{figure}
\begin{center}
\includegraphics[width=64mm]{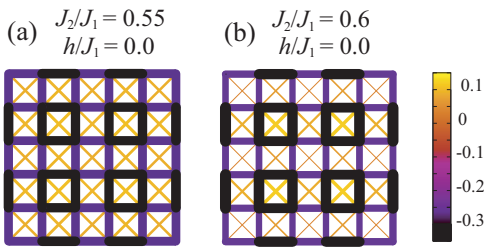}
\caption{(Color online) Nearest- and next-nearest-neighbor correlations $\langle\mathbf{S}_i\!\cdot\!\mathbf{S}_j\rangle$ of central 6 $\times$ 6 sites in the p-12 cluster at $h$ = 0. The thickness and color of the lines represent the magnitude and sign of the spin correlation $\langle\mathbf{S}_i\!\cdot\!\mathbf{S}_j\rangle$, respectively.
\label{pvbssou}}
\end{center}
\end{figure}
\begin{figure}
\begin{center}
\includegraphics[width=86mm]{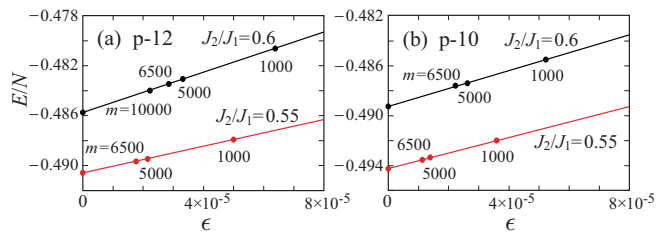}
\caption{(Color online) Ground-state energy per site, $E/N$, as a function of the truncation error, $\epsilon$, obtained by the DMRG method with the SSD in the p-12 (a) and p-10 (b) clusters at zero magnetic field. The number of kept states in the DMRG is $m$. The black and red solid lines show linear extrapolations for $J_2/J_1$ = 0.6 and $J_2/J_1$ = 0.55, respectively. 
\label{eps}}
\end{center}
\end{figure}

\subsection{PVBC in zero magnetic field}
\begin{figure}
\begin{center}
\includegraphics[width=86mm]{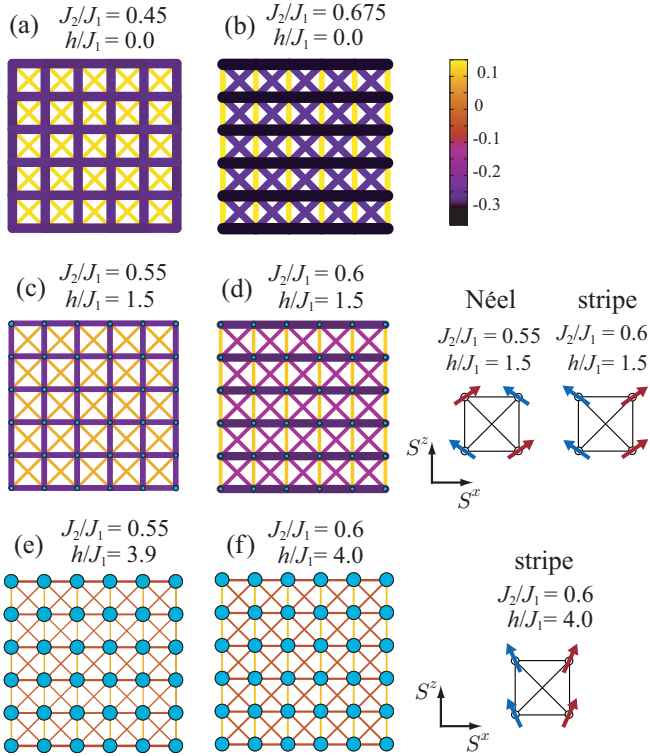}
\caption{(Color online) (a) and (b) $\langle\mathbf{S}_i\!\cdot\!\mathbf{S}_j\rangle$ of p-12 cluster and (c) -- (f) $\langle S^+_iS^-_j\rangle$ and $\langle S^z_i\rangle$ of p-12 or p-14 clusters. Only 6 $\times$ 6 sites in the center area are shown. The thickness and color of the lines represent the magnitude and sign of the spin correlations $\langle\mathbf{S}_i\!\cdot\!\mathbf{S}_j\rangle$ or $\langle S^+_iS^-_j\rangle$, respectively. The diameter of the circles on the lattice represents the magnitude of $\langle S^z_i\rangle$. The schematic figures on the right side with four arrows show the corresponding classical spin structure of each quantum spin state.
\label{NS}}
\end{center}
\end{figure}

Figure~\ref{pvbssou} shows the nearest- and next-nearest-neighbor correlation functions $\langle\mathbf{S}_i\cdot\mathbf{S}_j\rangle$ of 6 $\times$ 6 sites in the p-12 cluster at $h$ = 0, where the spin correlations of the 4-spin plaquette state can be clearly seen. This is a characteristic feature of the PVBC.
For comparison, we show the results at $J_2/J_1$ = 0.45 and 0.675 in Figs.~\ref{NS}(a) and \ref{NS}(b), in which the N$\acute{\rm e}$el and stripe correlations are obtained, respectively. These results show that the PVBC is realized only in a highly frustrated region around $J_2/J_1 \approx$ 0.55. The singlet-triplet spin gap $\Delta_{\rm st}$ is estimated to be 0.5$J_1$ from the magnetization curves in Fig.~\ref{ssdM-H}. This value is larger than that reported in other papers \cite{DMRG1,DMRG2,other6}. The PVBC ground state is also obtained in the p-10 cluster in zero magnetic field. Thus, we conclude that the PVBC is the ground state at around $J_2/J_1 \approx$ 0.55.
The reliability of the present results is shown in Fig.~\ref{eps}, which shows the ground-state energy per site $E/N$ as a function of the truncation error $\epsilon$ of the DMRG.
In both cases of $J_2/J_1$ = 0.55 and 0.6, the difference in energy relative to the extrapolated value in the limit of $\epsilon \rightarrow 0$ is less than 0.4\% for $m$ $>$ 5000 states.  
This shows that the accuracy of the ground-state energy is controlled by the truncation error $\epsilon$.
However, $E/N$ still have a difference of more than 0.7\% between the p-10 and p-12 clusters.
This is also the case for the order parameter of the PVBC defined as ${\cal O}_{\rm PVBC}$= $\overline{\langle\mathbf{S}_i\!\cdot\!\mathbf{S}_j\rangle}_{\rm w} - \overline{\langle\mathbf{S}_i\!\cdot\!\mathbf{S}_j\rangle}_{\rm s}$, where $\overline{\langle\mathbf{S}_i\!\cdot\!\mathbf{S}_j\rangle}_{\rm w}$ and $\overline{\langle\mathbf{S}_i\!\cdot\!\mathbf{S}_j\rangle}_{\rm s}$ are the averages of the weaker and stronger nearest-neighbor correlations, respectively. ${\cal O}_{\rm PVBC}$ at $J_2/J_1$=0.55 is 0.0308 and 0.0179 for the p-10 and p-12 clusters, respectively, 
while it is 0.0287 and 0.0990 at $J_2/J_1$=0.6. 
We must therefore perform a careful analysis of larger systems to determine the precise value of the order parameter.

\begin{figure}
\begin{center}
\includegraphics[width=72mm]{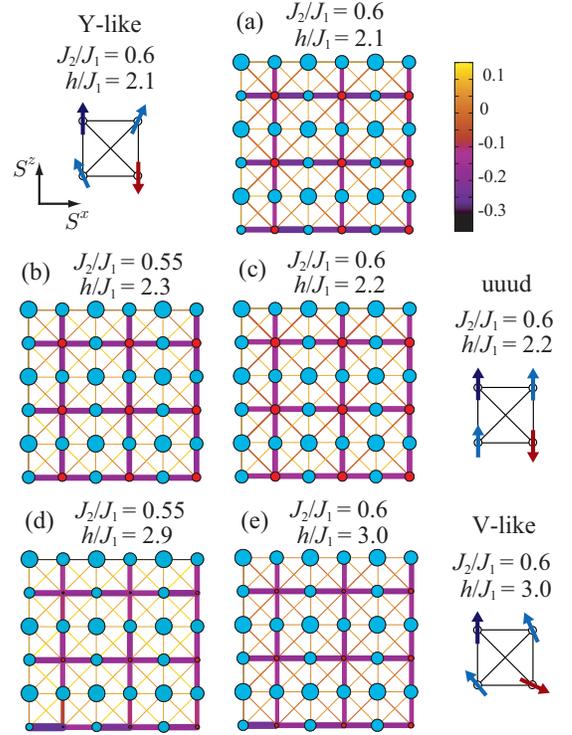}
\caption{(Color online) Distributions of $\langle S^+_iS^-_j\rangle$ and $\langle S^z_i\rangle$ of p-14 cluster in 1/2 plateau phase and the neighboring phases. The thickness and color of the lines represent the magnitude and sign of the spin correlations $\langle S^+_iS^-_j\rangle$, respectively. The diameter and the color of the circles on the lattice represent the magnitude of $\langle S^z_i\rangle$ and the sign (blue and red are positive and negative), respectively. The figures with four arrows show the schematic classical spin structures. 
\label{uuud}}
\end{center}
\end{figure}

\subsection{N$\acute{\mathbf{e}}$el and stripe phases in a magnetic field} 
The correlation functions of the nearest- and next-nearest-neighbor spins $\langle S^+_iS^-_j\rangle$ and the local magnetization $\langle S^z_i\rangle$ at finite magnetic fields  are shown in Figs.~\ref{NS}(c) - \ref{NS}(f).
In Fig.~\ref{NS}(c) we find strong antiferromagnetic correlation for nearest-neighbor  spins and ferromagnetic correlation for next-nearest-neighbor spins. These correlations are consistent with the N$\acute{\rm e}$el state shown in Fig.~\ref{NS}(a).
In Figs.~\ref{NS}(d), \ref{NS}(e) and \ref{NS}(f), the nearest-neighbor spin correlations are antiferromagnetic in horizontal directions but ferromagnetic in vertical directions, which is a feature of the stripe phase shown in Fig.~\ref{NS}(b).
The schematic classical analogues of these magnetic structures are represented in the right part of Fig.~\ref{NS}.
At $J_2/J_1$=0.55, we expect that the transverse N$\acute{\rm e}$el phase in a low magnetic field ($h/J_1<2.1$) will be stabilized by quantum fluctuations because the transverse N$\acute{\rm e}$el phase is not the ground state in the classical limit in this region. At a high magnetic field ($h/J_1 >$ 3.88), however, the quantum fluctuations are suppressed by the magnetic field and a stripe phase is realized with large uniform magnetization, as in the classical system.

\subsection{1/2 plateau phase and related states} 
We next discuss the results around the 1/2 plateau phase. As shown in Fig.~\ref{ssdM-H}(b), the magnetization at $J_2/J_1$=0.6 is continuous at the transition to the plateau phase. This result suggests a continuous change in the spin structure.
To understand how the spin structure is modified across the transition to the 1/2 plateau state, we show the correlation functions $\langle S^+_iS^-_j\rangle$ and the local magnetizations $\langle S^z_i\rangle$ at $J_2/J_1$ = 0.6 in Figs.~\ref{uuud}(a), \ref{uuud}(c) and \ref{uuud}(e).
Focusing on $\langle S^z_i\rangle$, we find four sublattice orders having an up-up-up-down structure with no distinct difference among these three phases. This kind of smooth change is also observed in the correlation functions $\langle S^+_iS^-_j\rangle$, which confirms a continuous transition. In these phases, one of the four spins is almost fully polarized (more than 95\% of full polarization) and the correlations with the nearest- and next-nearest-neighbor spins are negligibly small with absolute values of less than 0.05. Hence, the nearly fully polarized spins are almost independent of the other spins. The other three spins in the unit cell interact with each other through the antiferromagnetic interactions and they are in a similar situation to frustrated spins on a triangular lattice.
We therefore expect that the spin structures of these phases will be composed of one up-spin and the spin structure of the triangular lattice. We thus understand that the uuud state of the 1/2 plateau is the uud state of the 1/3 plateau of the triangular lattice \cite{tri1,tri2,tri3} with one additional up-spin.
Indeed, at $J_2/J_1$=0.5 the ground state of the Hamiltonian for the three spins is shown to be the exact ground state of the triangular Heisenberg model in the classical limit.
From the correspondence between the Hamiltonian for the three spins and the triangular Heisenberg model, the phases around the uuud phase should correspond to the Y phase and V phase of the triangular lattice \cite{tri1,tri2,tri3}.
We thus name the phases around $h/J_1 \sim$ 2.1 and 3.0 as the Y-like phase and V-like phase, respectively.
Although we have not yet confirmed the presence of an off-diagonal long-range order in the Y-like and V-like phases, the clear periodic structure of $\langle S^z_i\rangle$  and the stable transverse correlation $\langle S^+_i S^-_j\rangle$ with gapless spin excitations suggest that these phases have the characteristics of  a supersolid \cite{sps}.
The transition from the Y-like or V-like phase to the uuud phase is then expected to be a supersolid-solid transition.
The corresponding classical spin structures of the Y-like and V-like phases are represented in Fig.~\ref{uuud} with four arrows in the unit cell.
We note that such Y-like, V-like, and uuud states do not appear at $J_2/J_1$ = 0.45 and 0.675. 

\begin{figure}
\begin{center}
\includegraphics[width=66mm]{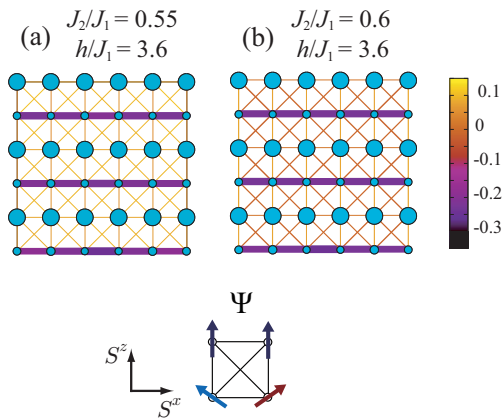}
\caption{(Color online) Nearest- and next-nearest-neighbor correlation functions $\langle S^+_iS^-_j\rangle$ and local magnetization $\langle S^z_i\rangle$ of p-14 cluster in $\Psi$ phase. The thickness and color of the lines represent the magnitude and sign of the correlations $\langle S^+_iS^-_j\rangle$, respectively. The diameter of the circles on the lattice represents the magnitude of $\langle S^z_i\rangle$. The bottom figure shows the schematic magnetic structure of the $\Psi$ phase. 
\label{PH}}
\end{center}
\end{figure}

\subsection{ $\Psi$ phase} 
In a high magnetic field of $h/J_1 \sim$ 3.6, close to the phase transition to the transverse stripe phase at $h/J_1 \sim$ 4, the one-dimensionally ordered state shown in Fig.~\ref{PH} is realized.
The distribution of the local magnetization $\langle S^z_i\rangle$ has a stripe structure mainly consisting of almost fully polarized spins.
The schematic classical spin structure is represented in the lower part of Fig.~\ref{PH},  and we name this phase the $\Psi$ phase.
We consider that with the reduction of the dimensionality, the quantum fluctuations are enhanced under competition with the Zeeman energy that stabilizes fully polarized spins.
Similarly to the one up-spin in the magnetic unit cell of the Y-like, uuud, and V-like phases, almost fully polarized spins are nearly independent of the other spins.
In the Y-like, uuud, and V-like phases, each of the almost fully polarized spins is surrounded by eight quantum mechanically correlated spins. Since the magnetic unit cell includes 4-spins, one out of four spins is ejected from the quantum mechanically correlated spins.
In the $\Psi$ phase, two out of four spins are ejected and one-dimensionally correlated spin chains are formed. 
These states have a common feature of forming domains of nearly independent fully polarized spins, which are expected to be realized in the highly frustrated region 
where the Zeeman energy competes with the quantum fluctuations.
In the periodic-boundary conditions (torus) at $J_2/J_1$=0.5, the stripe structure of the local magnetization has been rigorously proved to be an eigenstate \cite{uuud2,psi}.
We therefore expect that the $\Psi$ phase in a high magnetic field will be stable even at $J_2/J_1 \ne$ 0.5, although the phase boundary between the V-like phase and the $\Psi$ phase is still to be resolved.
This transition is expected to be second- or higher order since we have no signal of a first-order level crossing. Even if we use the SSD, critical long-range correlations are affected by the presence of a boundary, which will smear the critical behavior around the transition. We therefore require finite-size scaling analysis to clarify the nature of the transition. 
\begin{figure}
\begin{center}
\includegraphics[width=86mm]{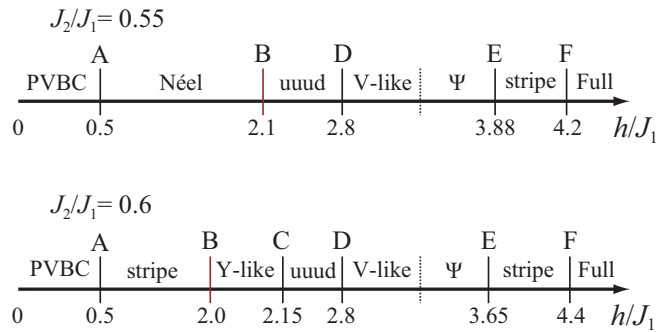}
\caption{(Color online) Ground-state phase diagram of $S=1/2$ $J_1$-$J_2$ Heisenberg model on the square lattice at $J_2/J_1$ =  0.55 (above) and 0.6 (bottom) in magnetic fields. The red lines show first-order transitions. Letters at the phase boundaries correspond to those in Fig.~\ref{ssdM-H} at the anomalies in the magnetization curve.
\label{phase}}
\end{center}
\end{figure}

\subsection{First-order transition }
Normally the spin-flop transition is a phenomenon involving the reorientation transition of spins from the easy axis of anisotropic systems. 
This kind of discontinuous change in the magnetic structure is a first-order transition. A similar transition may occur under the effect of frustration and quantum or thermal fluctuations even in the isotropic Heisenberg model without an easy axis, although it is not trivial to assign the classical spin orientation in quantum systems \cite{spinflop1,spinflop2}.
In the present study, we find a jump in magnetization at the phase transition from the transverse N$\acute{\rm e}$el- and stripe-ordered phases to the uuud and Y-like phases, respectively, labeled by B in Fig.~\ref{ssdM-H}. Across the transition, the spin correlation functions entirely change, and we consider that this transition is a kind of first-order spin-flop transition from a transverse N$\acute{\rm e}$el- or stripe-ordered state to a uuud or Y-like ordered state occurring in the isotropic Heisenberg model. In contrast to the normal spin flop from the parallel direction to the perpendicular direction to the magnetic field, this transition is a spin flop from the perpendicular direction to the parallel direction  to the magnetic field.

\section{ Summary }
We have investigated the magnetization process of the $S=1/2$ $J_1$-$J_2$ Heisenberg model on the square lattice at zero temperature by the DMRG + SSD method.
The obtained ground-state phase diagram consists of eight different phases, as shown in Fig.~\ref{phase}, including a $\Psi$ phase characterized by a stripe structure in a high magnetic field, and Y-like and V-like phases around the uuud 1/2 plateau phase, whose spin structures are continuously connected with each other. The $\Psi$, V-like, uuud, Y-like, and PVBC states are found only near the highly frustrated region around $J_2/J_1 \sim$ 0.55.
Since these phases are not realized in the ground state in the classical limit, the appearance of such phases is a result of the strong frustration and quantum fluctuation of the $S=1/2$ $J_1$-$J_2$ Heisenberg model.

\section*{Acknowledgment} The present work was supported by a Grant-in-Aid for
Scientific Research (No. 26400344) from JSPS.

\end{document}